\documentclass{webofc}
\usepackage[varg]{txfonts}   
\usepackage{paralist}

\begin{document}
\title{GPU-based Online Track Reconstruction for the ALICE TPC in Run~3 with Continuous Read-Out}
%
%

\author{
  \firstname{David} \lastname{Rohr}\inst{1}\fnsep\thanks{\email{drohr@cern.ch}}
  \and \firstname{Sergey} \lastname{Gorbunov}\inst{2,3}
  \and \firstname{Marten Ole} \lastname{Schmidt}\inst{4}
  \and \firstname{Ruben} \lastname{Shahoyan}\inst{1}
}

\institute{European Organization for Nuclear Research (CERN), Geneva, Switzerland
\and
           Frankfurt Institute for Advanced Studies, Ruth-Moufang-Str.~1, 60438 Frankfurt, Germany
\and
           Goethe University Frankfurt, Germany
\and
           University of Heidelberg, Germany
}

\abstract{%
    In LHC Run~3, ALICE will increase the data taking rate significantly to 50 kHz continuous read-out of minimum bias Pb--Pb collisions.
    The reconstruction strategy of the online-offline computing upgrade foresees a first synchronous online reconstruction stage during data taking enabling detector calibration and data compression, and a posterior calibrated asynchronous reconstruction stage.
    Many new challenges arise, among them continuous TPC read-out, more overlapping collisions, no a priori knowledge of the primary vertex and of location-dependent calibration in the synchronous phase, identification of low-momentum looping tracks, and sophisticated raw data compression.
    The tracking algorithm for the Time Projection Chamber (TPC) will be based on a Cellular Automaton and the Kalman filter.
    The reconstruction shall run online, processing 50 times more collisions per second than today, while yielding results comparable to current offline reconstruction.
    Our TPC track finding leverages the potential of hardware accelerators via the OpenCL and CUDA APIs in a shared source code for CPUs and GPUs for both reconstruction stages.
    We give an overview of the status of Run~3 tracking including performance on processors and GPUs and achieved compression ratios.
}
\maketitle
\section{Introduction}

ALICE (A Large Ion Collider Experiment \cite{bib:alice}) is one of the four major experiments at the LHC (Large Hadron Collider) at CERN.
It is a dedicated heavy-ion experiment studying lead collisions at the LHC at unprecedented energies.
During the second long LHC shutdown in 2019 and 2020, the LHC upgrade will provide a higher Pb--Pb collision rate, and ALICE will update many of its detectors and systems~\cite{bib:aliceupgrade}.
In particular, the main tracking detectors TPC (Time Projection Chamber) and ITS (Inner Tracking System) will be upgraded~\cite{bib:tpcrun3tdr}, and the computing scheme will change with the O$^2$ online-offline computing upgrade~\cite{bib:o2tdr}.

ALICE will upgrade the detectors for LHC Run~3 and switch from the current triggered read-out of up to 1\,kHz of Pb--Pb events to a continuous read-out of 50\,kHz minimum bias Pb--Pb events.
The continuous read-out of pp collisions will happen at rates between~200\,kHz and~1\,MHz.
ALICE is abandoning the hardware triggers and will switch to a full online processing in software.
During data taking, the synchronous processing will serve two main objectives: detector calibration and data compression.
With a flat budget and the yearly increases of storage capacity, recording and storing raw data as today is prohibitively expensive at 50 to 100 times the data rate.
ALICE aims at a compression of the TPC data, the largest contributor to raw data size, of a factor 20 compared to the zero-suppressed raw data size of Run 2.
By producing the calibration during data taking, ALICE will reduce the number of offline reconstruction passes over the data, where the first two passes serve the calibration today.
After the data taking, the asynchronous reconstruction will reprocess the data for the final reconstruction output.
This asynchronous stage will employ the same algorithms and software as the synchronous stage, but with different settings, additional reconstruction steps, and final calibration.

\section{Tracking}

The most time consuming step during event processing is the reconstruction of particle trajectories, which involves mostly 3 detectors in the case of ALICE: the TPC, the ITS, and the TRD (Transition Radiation Detector).
In order to leverage the potential of modern parallel hardware accelerators, ALICE is designing tracking implementations that shall run on GPU (Graphics Processing Units), which offer significantly more compute power for parallel applications compared to traditional processors.
Due to the continuous read-out of the TPC without trigger, hits in the TPC do not have a defined $z$-position but only a time value.
The $z$-position, and also the calibration for the hit, depend on the time of the interaction that produced this hit, which is ambiguous in the continuous read-out.
ALICE has adjusted the tracking algorithm accordingly~\cite{bib:ctd2018}:
\begin{compactitem}
\item In the seeding phase of TPC tracking, the tracking does not consider the calibration at all, but simply scales the time to $z$ linearly using the average drift velocity, and assumes that all tracks are primary tracks originating from the origin;
\item This allows a first preliminary fit, which is used to find the full track, improve the $z$-estimation, and finally apply the calibration, still under the assumption of a primary track;
\item The tracking in the silicon detector ITS is not sensitive to calibration, and ITS tracks are reconstructed standalone in parallel;
\item Tracks reconstructed in TPC and ITS are matched in the compatible time window, which finally fixes the time and thus the $z$-position of the TPC tracks;
\item This enables the final refit of the track in the TPC and the propagation to the TRD layers outside the TPC.
\end{compactitem}
\section{TPC Tracking}

Due to the vast number of hits in the TPC, TPC tracking is computationally expensive.
For other detectors, the synchronous processing can be limited to a subset of the events needed for calibration, but the TPC needs the full set of events for the data compression (see section~\ref{sec:comp}).
This makes the TPC tracking the critical part of online computing.
The TPC tracking algorithm has been derived from the current ALICE HLT (High Level Trigger) TPC tracking \cite{bib:tns,bib:chep,bib:cnna,bib:chep2016gpu}.
The code is written in a generic vendor- and API-independent way that supports GPUs via the CUDA and OpenCL languages, and parallelizes over CPU cores via OpenMP~\cite{bib:generic}.

\begin{figure}[htb]
\centering
\includegraphics[width=0.9\textwidth]{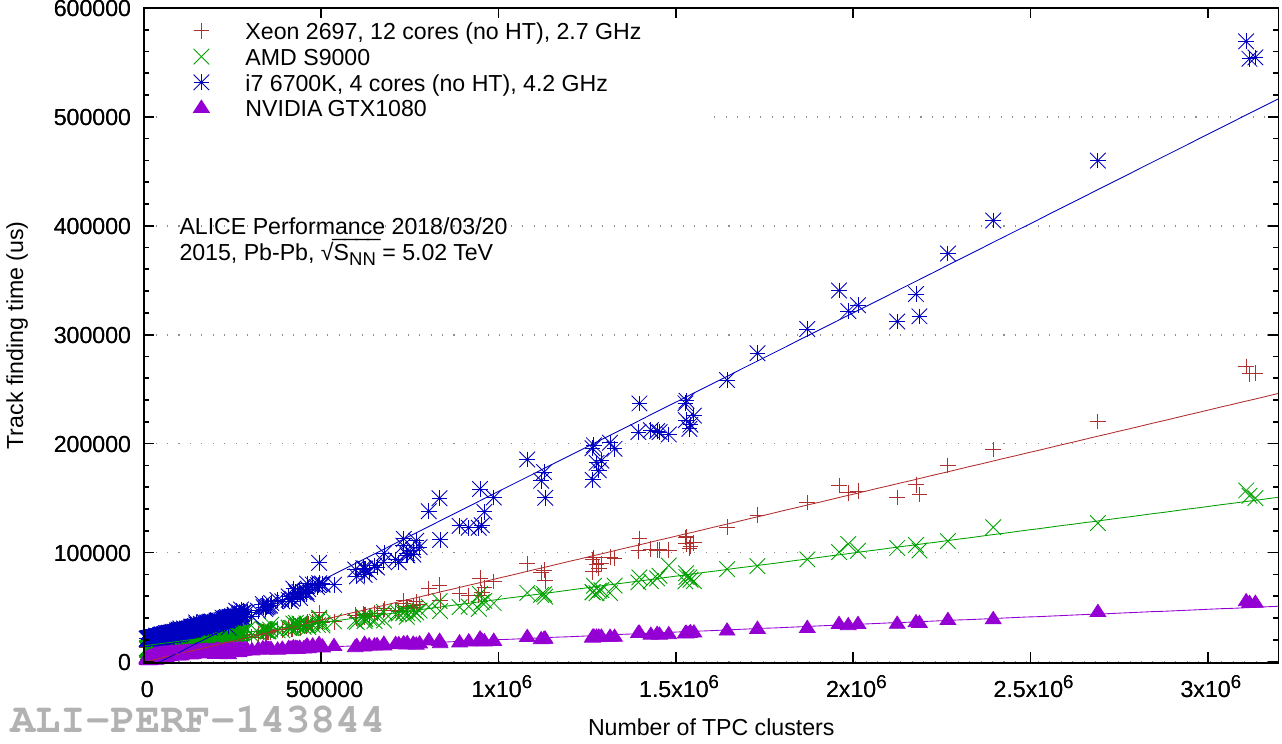}
\caption{Run time of the ALICE HLT TPC Tracking on CPU and GPU versus event size}
\label{fig:speed}
\end{figure}

Principally, the combinatorial complexity for combining hits into tracks goes much faster than linearly with the number of hits.
With the large number of hits during continuous data taking of Pb--Pb events, any dependency that is more than linear is unacceptable with the available compute resources.
The TPC tracking handles the combinatorial part in the early seeding phase using a Cellular Automaton~\cite{bib:tns}.
Due to the abundance of hits in the TPC, it is not necessary to follow multiple track hypotheses.
Therefore, all posterior steps after the seeding have a linear run time with respect to the event size.
Figure~\ref{fig:speed} demonstrates that the tracking duration is almost linear both on CPU and on GPU, irrespective of the processor model.

\begin{figure}[htb]
\centering
\includegraphics[width=0.9\textwidth]{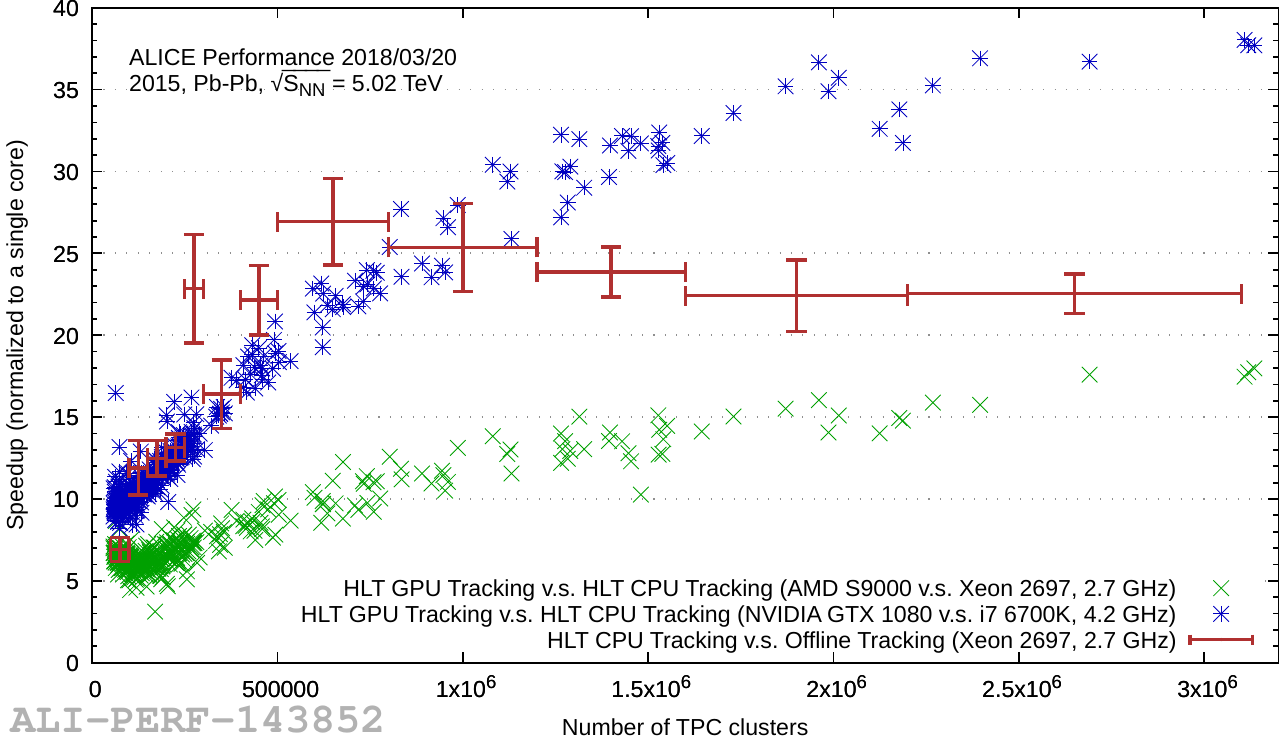}
\caption{Speedup of the ALICE TPC tracking on GPU normalized to a single CPU core}
\label{fig:speedup}
\end{figure}

The evaluation of the speedup of the GPU version over the CPU version needs to take into account that the GPU version needs certain CPU resources for pre- and post-processing.
Since the number of CPU cores used for tracking is a free parameter, we compute the GPU speedup versus a single core.
The tracking scales pretty much linearly with the number of CPU cores, and the speedup versus more cores is accordingly smaller.
We need to correct for the CPU contribution to the GPU version, since also the GPU version consumes certain CPU resources.
Therefore, we subtract the CPU runtime\footnote{CPU time as measured by Linux, i.\,e.~sum of the actual CPU time of all threads without the idle periods.} during the GPU version from the CPU runtime of the CPU version.
In other words, we compute how many CPU cores are equivalent to a GPU, which of course depends on the GPU and CPU model.
Figure~\ref{fig:speedup} shows that with the hardware installed in the current HLT farm, the GPU can replace up to 18 CPU cores for large events, while more recent GPUs can replace up to 40 CPU cores.
The figure contains the profile for the speedup of HLT tracking versus offline tracking in red.
The speedup of a GPU versus a CPU core running the offline tracking is therefore the product, which yields a speedup of up to 800 with a modern GPU and Pb--Pb events versus a single core running the current offline tracking.

The current HLT tracking is comparable with the ALICE offline tracking in terms of efficiency but yields a worse resolution due to simplifications in the track fit~\cite{bib:lhcp2017}.
For Run~3, many improvements have been performed and several features of offline tracking, like cluster rejection, 3-dimensional magnetic field, multiple fit iterations, etc., have been ported.
The current development version of the Run~3 tracking yields the same resolution as offline on a global view~\cite{bib:ctd2018}.
Locally, e.\,g.~at the TPC sector boundaries, there are still some improvements to be done.

In order to obtain the current TPC drift speed, an online TPC drift velocity calibration has been implemented in the HLT~\cite{bib:chep15,bib:ctd,bib:tns2016}, and a similar feature is foreseen for the synchronous stage of the Run~3 tracking.

\section{TPC to ITS matching}

\begin{figure}[htb]
\centering
\includegraphics[width=9cm,clip]{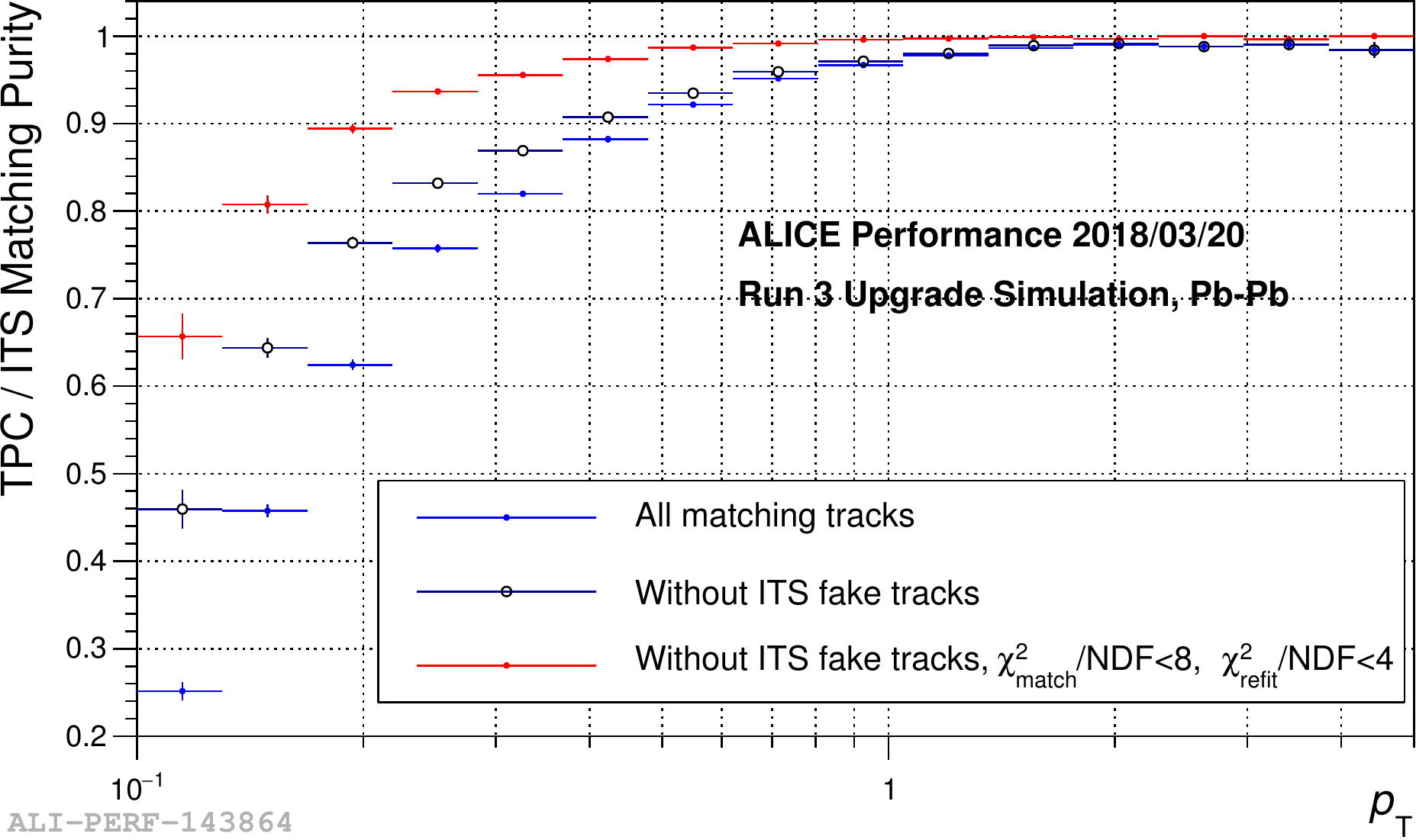}
\caption{Matching purity of TPC to ITS tracks in current development version of the Run~3 reconstruction software versus track transverse momentum ($p_{\text{T}}$)}
\label{fig:itsmatching}
\end{figure}

A standalone ITS tracking implementation is currently being developed for Run~3, which will support GPUs and use a Cellular Automaton.
Meanwhile, a fast ad-hoc ITS tracking is used for performance studies, which was implemented in a short time and does not achieve full efficiency.
Figure~\ref{fig:itsmatching} shows the purity of the TPC to ITS matching versus transverse momentum.
The ad-hoc ITS tracking is not optimized and yields many fake tracks.
Eventually, it will be replaced by the Cellular Automaton based tracking, but for the performance studies we use Monte-Carlo labels to remove fake tracks (black and red curves).
If we apply strict $\chi^2$ cuts for ITS, the matching is as good as in the current Run 2.
This is another demonstration that the Run~3 TPC tracking works well.

Due to the high track density in the inner layers, and due to the necessity to follow multiple track hypotheses, ITS standalone tracking is combinatorially more complex than TPC tracking.
However, for calibration in the synchronous phase, it will be sufficient to process only a subset of the events, and employ a simplified and faster version of the tracking, which finds only primary tracks without missing layers and at relatively high $p_{\text{T}}$ of at least~500~MeV/$c$.

\section{TPC to TRD prolongation tracking}

Today, both hits in the TRD as well as on-the-fly reconstructed TRD online tracklets are stored on tape.
The hits are used for the offline tracking, while the online tracklets are used to generate trigger decisions.
Due to bandwidth constraints, the TRD read out in Run 3 is restricted to the tracklets, such that the tracking implementation must change to use tracklets instead of hits.

In contrast to the ITS, ALICE does currently not foresee to perform standalone TRD tracking.
ITS standalone tracking is mandatory, since prolonging TPC tracks into the ITS is infeasible due to the lack of an absolute $z$-coordinate in the TPC before matching with ITS.
Because of a high fraction of fake TRD online tracklets, TRD standalone tracking would be computationally too expensive.
Instead, the TRD tracking works by prolonging TPC tracks into the TRD after their $z$-coordinate was fixed by the TPC ITS matching.

\begin{figure}[htb]
 \begin{minipage}[t]{0.485\textwidth}
 \centerline{\includegraphics[width=0.955\textwidth]{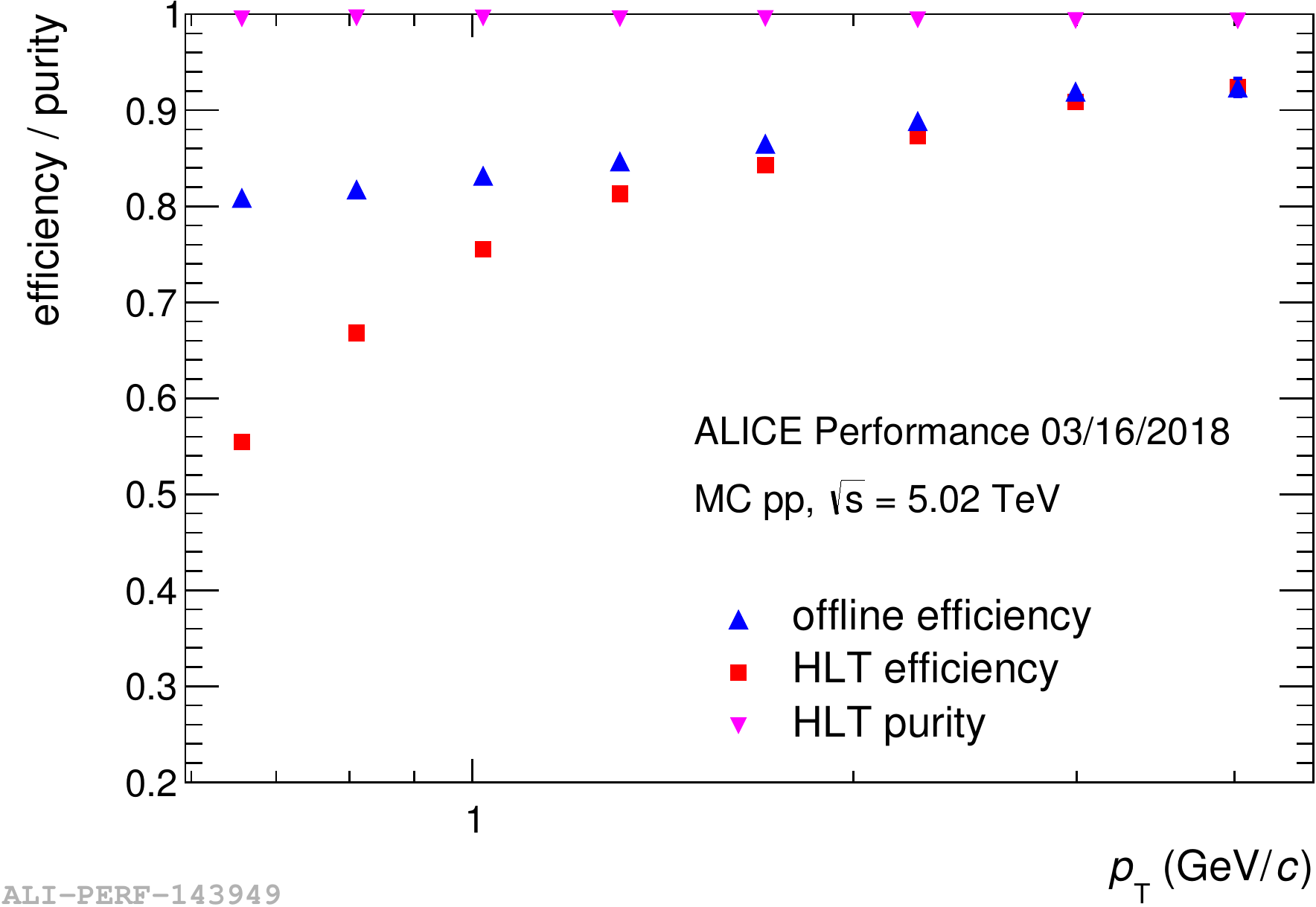}}
 \caption{TRD tracking efficiency and purity in offline reconstruction and in HLT with pp data}
 \label{fig:trdpp}
 \end{minipage}
 \hfill
 \begin{minipage}[t]{0.485\textwidth}
 \centerline{\includegraphics[width=0.955\textwidth]{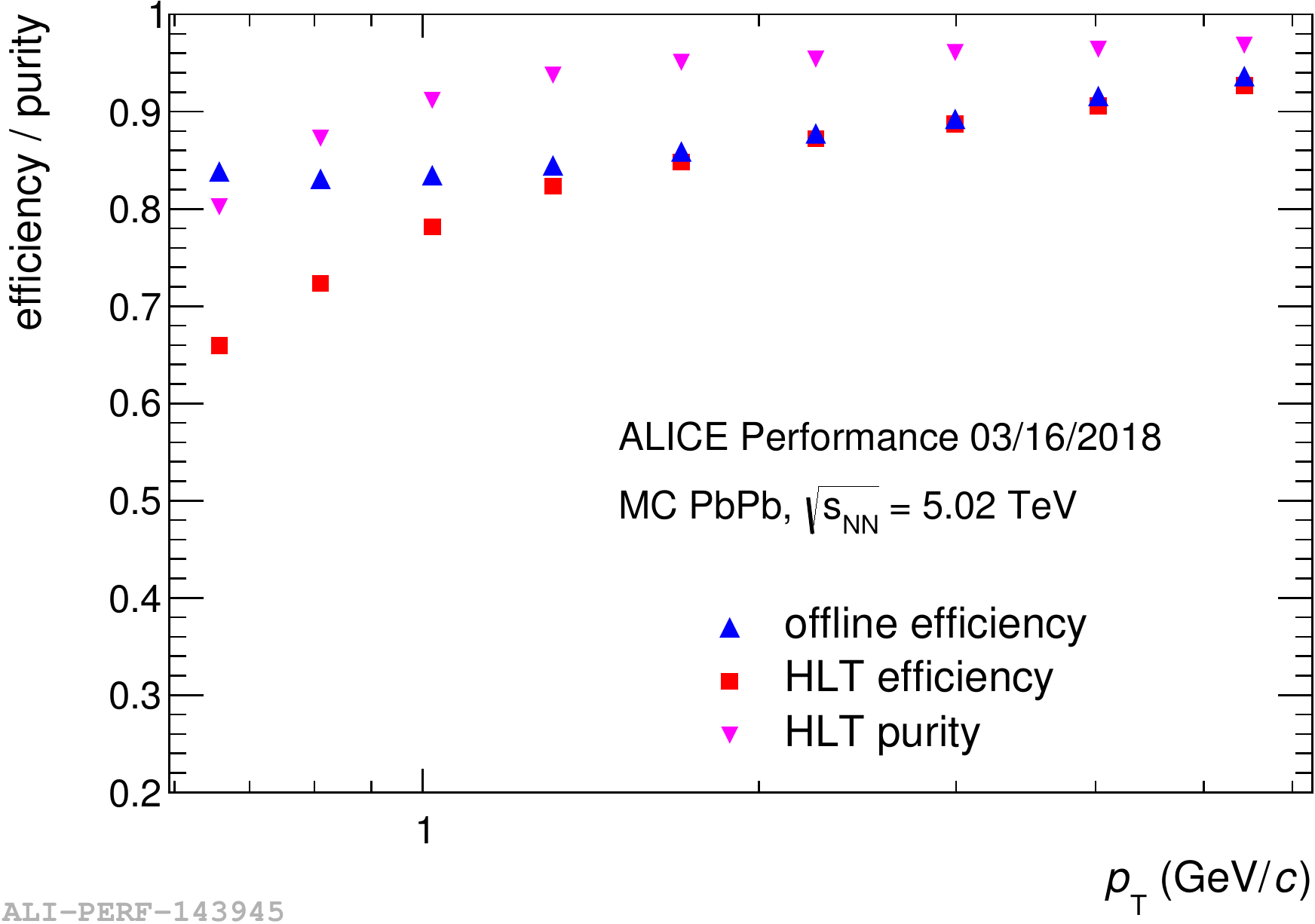}}
 \caption{TRD tracking efficiency and purity in offline reconstruction and in HLT with Pb--Pb data}
 \label{fig:trdpbpb}
 \end{minipage}
\end{figure}

The new TRD tracking has been developed and is tested in the ALICE HLT.
Figures~\ref{fig:trdpp} and \ref{fig:trdpbpb} compare the TRD tracking efficiency and the purity of the new online tracklet-based tracking in the HLT with the offline tracking.
A track is considered found if the TPC track has a prolongation into the TRD with matched online tracklets in at least 2 TRD layers.
While the efficiency at high-$p_{\text{T}}$ is comparable, the offline tracking yields a higher efficiency at low-$p_{\text{T}}$.
The reason are the online tracklets, which not yet optimized for low-$p_{\text{T}}$.
Therefore, a large fraction of low-$p_{\text{T}}$ tracklets is simply missing.
This is a general deficiency of the approach, but there are ongoing studies to extend the online tracklet finding towards lower-$p_{\text{T}}$.
While the purity is perfect for pp data, it goes down to 80\% for low-$p_{\text{T}}$ Pb--Pb data, due to the high density of tracklets in the TRD.

\section{TPC Data Compression}
\label{sec:comp}

\begin{figure}
\centering
\includegraphics[width=7cm,clip]{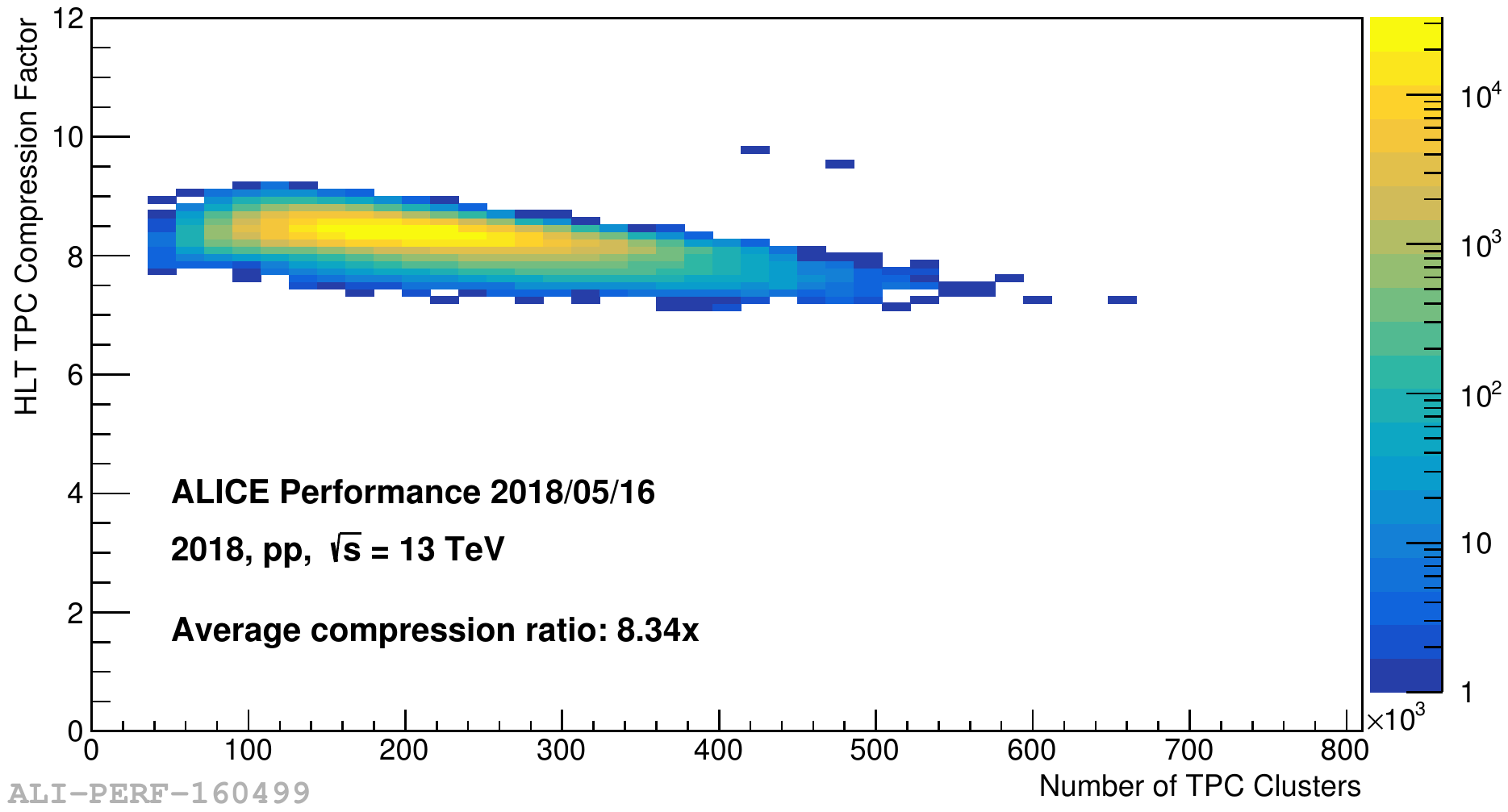}
\caption{Compression factor of ALICE TPC data in the HLT in 2018 with pp data}
\label{fig:comp}
\end{figure}

The most important task of the synchronous reconstruction is the compression of the raw data, in order to enable the storage of all data.
We discuss this topic for the TPC, the largest contributor to data size.
The compression involves several steps:
\begin{compactenum}
 \item Replacement of raw data by TPC hits reconstructed in real time. This step itself has only a moderate compression effect, but it is required for the following steps.
 \item The clusters are converted to an integer format, storing only as many bits as required with respect to the intrinsic TPC resolution. Cluster shape and charge are encoded with a dynamic precision relative to their absolute value.
 \item The entropy of the cluster properties is reduced, mainly by storing residuals of clusters to tracks for clusters attached to tracks, and position differences otherwise~\cite{bib:lhcp2017}.
 \item Clusters are compressed with a standard entropy compressor. We are investigating Huffman encoding, Arithmetic encoding, or ANS encoding.
 \item Clusters not needed for physics analysis are removed.
\end{compactenum}
Steps 1 to 4 are implemented in the current HLT, yielding a compression factor of~8.34 in pp events, as shown in Fig.~\ref{fig:comp}.
The prototype for the Run~3 TPC data compression uses more elaborate versions of the algorithms and achieves a compression factor of~9.1 for Pb--Pb events.

\begin{figure}[htb]
\centering
\includegraphics[width=\textwidth]{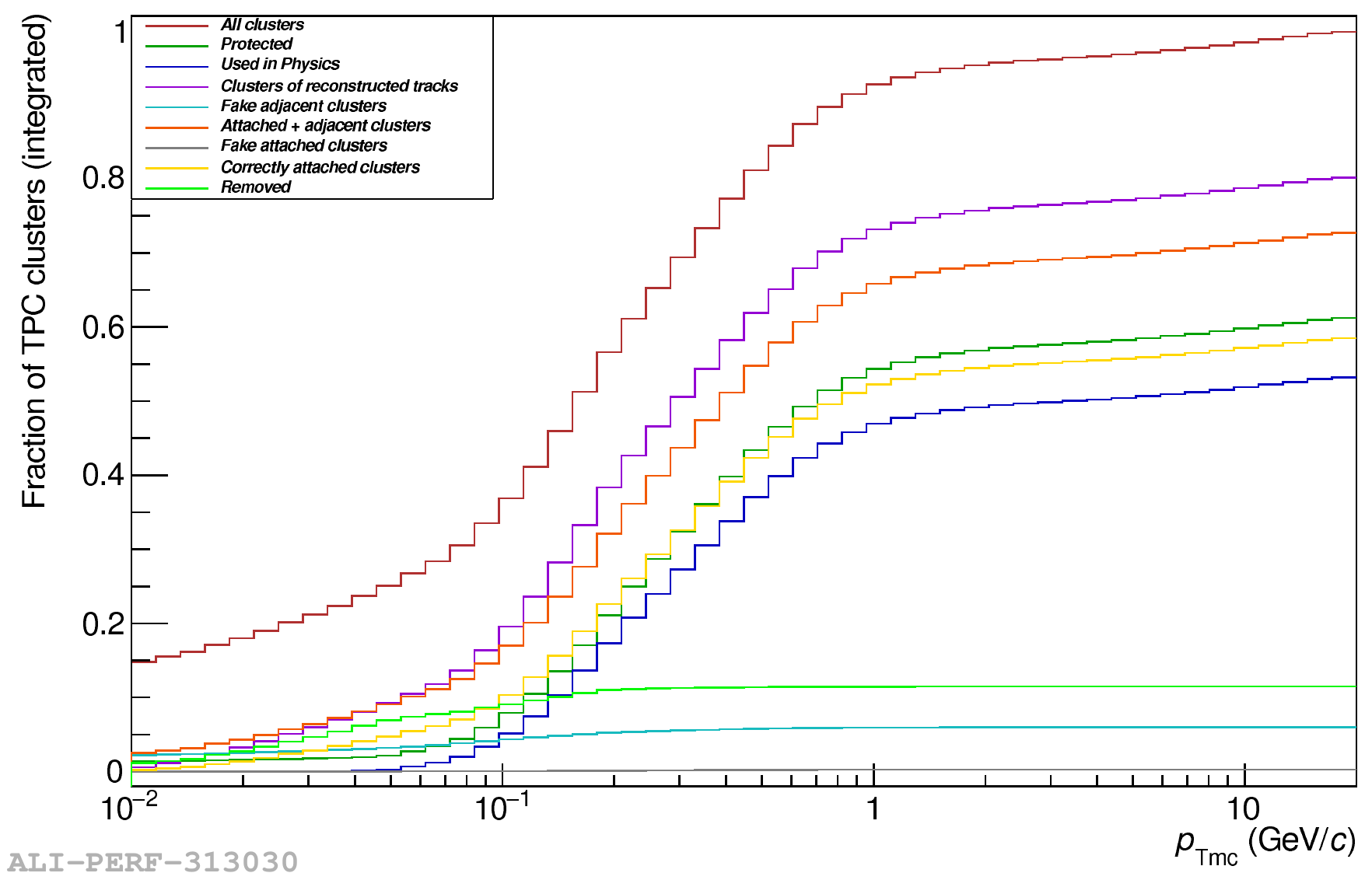}
\caption{Integrated cluster attachment and rejection distribution of the ALICE TPC with the current development version of the Run~3 reconstruction software}
\label{fig:clusters}
\end{figure}

For Run~3, ALICE will implement an additional step that rejects clusters not used for physics.
Figure~\ref{fig:clusters} illustrates the cluster attachment status of the current development version of the tracking.
We aim to remove the following types of clusters:
\begin{compactitem}
 \item Clusters of tracks below~50~MeV/$c$.
 \item Clusters of secondary legs of looping tracks below~200~MeV/$c$.
 \item Clusters of track segments with high inclination angle that are not used in the track fit.
 \item Clusters from noise and from broken TPC pads.
\end{compactitem}

The figure shows the integrated fraction of clusters with certain properties versus transverse momentum starting at~10~MeV/$c$, normalized to all clusters.
The curve for all clusters starts at 14\%, which means that 14\% of the clusters stem from noise or from tracks below~10~MeV/$c$.
These clusters could safely be removed but are not accessible via tracking.
Therefore, ALICE is investigating other means to remove them.

Attached clusters in Figure~7 are those attached to tracks in the tracking and used for the fit.
The tracker forms tubes of 1.5 cm around each track, and marks unattached clusters as adjacent to the tube of the track with the highest transverse momentum it lies in.
Clusters marked as used in physics are those attached to good physics tracks, and not belonging to the above categories of clusters that shell be removed.
The clusters marked as protected lie in the 1.5 cm tube around the good track segments that are used in physics.
Removed clusters are all attached or adjacent clusters that are not protected.
Fake attached or adjacent clusters are those not carrying a Monte-Carlo label of the track they are attached to.
The figure clearly shows that the fake attachment, which is used in physics, is practically zero, while the adjacent clusters have a moderate fake rate.
The latter is due to the fact that they lie in the tube around a random high-$p_{\text{T}}$ track and are thus marked as adjacent to the wrong track.
This does not affect the tracking at all, but leads to a slightly lower cluster rejection rate.
Currently, the tracking is capable of removing around 12\% of the hits, another 14\% of clusters of tracks below~10~MeV/$c$ should be removed by another method.
The false positive rate of the track-based cluster rejection is in the order of one per mille and thus negligible.
Eventually, the detection of clusters in the tube will be tuned, and the $\chi^2$ should be considered instead of relying on a fixed tube of 1.5 cm radius.

\section{Conclusions}

ALICE will perform a major upgrade of the tracking detectors and for the computing for the LHC Run~3.
New tracking implementations are developed for the TPC, ITS, and TRD detectors, that will run on GPUs to speed up the processing, using a generic programming approach to ensure vendor independence.
The TPC tracking is implemented with comparable efficiency and resolution as in Run 2 offline tracking.
There are ongoing studies about how the tracking will behave with respect to the modified calibration procedure foreseen for Run~3..
First tests of the TRD tracking show comparable performance as in Run 2 offline tracking except for low-$p_{\text{T}}$ tracks, which is a known deficit of the Run~3 approach.
A critical aspect is the compression of the TPC data.
The current compression algorithm based on cluster finding an entropy encoding yields a compression factor of 9.1 in the Run~3 prototype.
In addition, the tracking allows for the removal of 12\% of hits which are certainly not used in physics.
Improvements in the tracking have the potential to increase this rejection ration, and another method will be needed to reject clusters belonging to tracks below~10~MeV/$c$.


\begin{thebibliography}{}

\bibitem{bib:alice}
{ALICE Collaboration},
Journal of Instrumentation {\bf 3} S08002 (2008)

\bibitem{bib:aliceupgrade}
{ALICE Collaboration},
``{Upgrade of the ALICE Experiment: Letter of Intent}'',
CERN-LHCC-2012-012 (2012)

\bibitem{bib:tpcrun3tdr}
{ALICE Collaboration},
``Technical Design Report for the Upgrade of the ALICE Time Projection Chamber'',
CERN-LHCC-2013-020 (2013)

\bibitem{bib:o2tdr}
{ALICE Collaboration},
``Technical Design Report for the Upgrade of the Online-Offline Computing System'',
CERN-LHCC-2015-006, ALICE-TDR-019 (2015)

\bibitem{bib:ctd2018}
D.~Rohr {\it et al.},
``Track Reconstruction in the ALICE TPC using GPUs for LHC Run 3'',
Presented at the 4th International Workshop Connecting the Dots (2018)
arXiv:1811.11481

\bibitem{bib:tns}
S.~Gorbunov {\it et al.},
IEEE Transactions on Nuclear Science {\bf 58}, 1845 (2011)

\bibitem{bib:chep}
D.~Rohr {\it et al.},
J.~Phys.: Conf.~Series {\bf 396}, 012044 (2012)

\bibitem{bib:cnna}
D.~Rohr,
``ALICE TPC Online Tracker on GPUs for Heavy-Ion Events''
Proceedings of 13th International Workshop on Cellular Nanoscale Networks and their Applications, pp. 1-6 (2012)

\bibitem{bib:chep2016gpu}
D.~Rohr {\it et al.},
J.~Phys.: Conf.~Series {\bf 898}, 32030 (2017)

\bibitem{bib:generic}
D.~Rohr {\it et al.},
``Portable and Vendor-Independent Low-Level Programming and Performance Benchmarking for Graphics Cards and Processors'',
2017 IEEE 19th International Conference on High Performance Computing and Communications Workshops, pp. 1-8 (2017)

\bibitem{bib:lhcp2017}
D.~Rohr for the {ALICE} Collaboration,
``{Tracking performance in high multiplicities environment at ALICE}'',
Presented at 5th Large Hadron Collider Physics Conference (2017), arXiv:1709.00618

\bibitem{bib:chep15}
D.~Rohr {\it et al.},
J.~Phys.: Conf.~Series {\bf 664}, 082047 (2015)

\bibitem{bib:ctd}
D.~Rohr {\it et al.},
EPJ Web Conf. {\bf 127}, 00014 (2016)

\bibitem{bib:tns2016}
D.~Rohr {\it et al.},
IEEE Transactions on Nuclear Science {\bf 64}, 1263 (2017)

\end{thebibliography}
\end{document}